\documentclass[conference]{IEEEtran}
\usepackage{fancyhdr}
\fancypagestyle{IEEEtitlepagestyle}{
  \fancyhf{}
  \fancyfoot[C]{\sffamily\footnotesize\thepage}
  \fancyfoot[L]{\sffamily\scriptsize Copyright held by the owner/author(s). Presented at\\
3rd International Workshop on Overlay Architectures for FPGAs (OLAF2017)
\\Monterey, CA, USA, Feb. 22, 2017.}

}
\usepackage{cite}
\usepackage{graphicx}
\usepackage{booktabs, multirow}

\begin{document}

\title{A Scalable, Low-Overhead Finite-State Machine Overlay for Rapid FPGA Application Development}

\author{\IEEEauthorblockN{David Wilson, Greg Stitt}
\IEEEauthorblockA{Department of Electrical and Computer Engineering\\
University of Florida\\
Gainesville, FL 32611\\
Email: d.wilson@ufl.edu, gstitt@ece.ufl.edu}}

\maketitle

\begin{abstract}
Productivity issues such as lengthy compilation and limited code reuse have restricted usage of  field-programmable gate arrays (FPGAs), despite significant technical advantages. Recent work into overlays---virtual coarse-grained architectures implemented atop FPGAs---has aimed to address these concerns through abstraction, but have mostly focused on pipelined applications with minimal control requirements. Although research has introduced overlays for finite-state machines, those architectures suffer from limited scalability and flexibility, which we address with a new overlay architecture using memory decomposition on transitional logic. Although our overlay provides modest average improvements of 15\% to 29\% fewer lookup tables for individual finite-state machines, for the more common usage of an overlay supporting different finite-state machines, our overlay achieves a 77\% to 99\% reduction in lookup tables. In addition, our overlay reduces compilation time to tenths of a second to enable rapid iterative-development methodologies.
\end{abstract}

\IEEEpeerreviewmaketitle

\section{Introduction} \label{sec:intro}
In various application domains, field-programmable gate arrays (FPGAs) have significant performance and energy advantages over other devices \cite{trimberger:three}, but often go unused due to low application-design productivity. For example, depending on the device and application, FPGA compilation can take on the order of hours or days \cite{lavin:hmflow}. Similarly, FPGA applications are generally unable to leverage large amounts of existing code, due to the need for device-specific code \cite{kirchgessner:virtualrc}, or due to unattractive tradeoffs in IP core libraries \cite{wilson:unified}.

\begin{figure}[!ht]
\centering
\includegraphics[width=0.85\columnwidth]{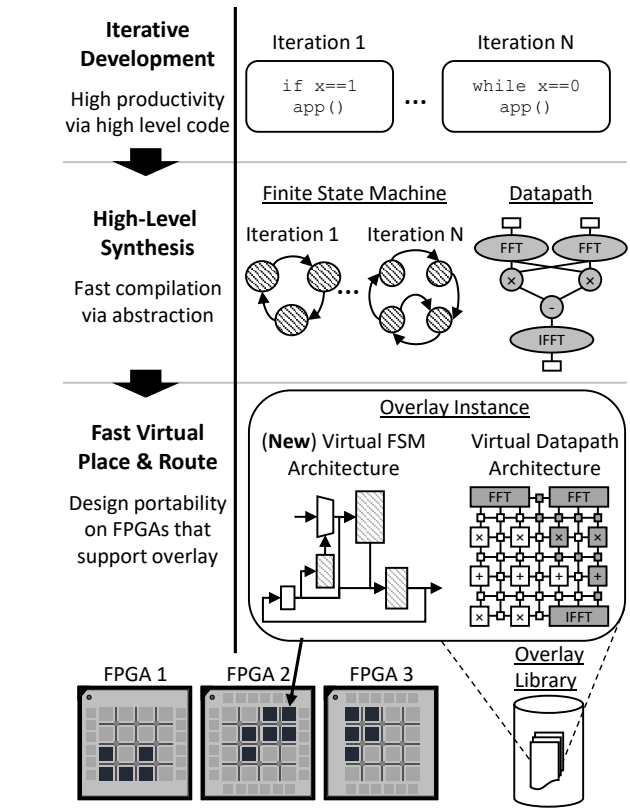}   
\caption{High-level synthesis using overlays to enable fast compilation, portability, and rapid iterative-design methodologies}
\label{fig:overview}
\end{figure}

Research in coarse-grained virtual architectures (overlays) has focused on overcoming these productivity limitations by providing an abstraction over the FPGA's fine-grained resources. By mapping design logic onto application-specialized resources, an overlay can achieve both portability and orders of magnitude faster compilation by avoiding decomposition of logic into thousands of FPGA lookup tables. Recent overlay research has shown place-and-route speedup of over 10,000$\times$ compared to vendor tools  \cite{coole:adjustable}. 

One key limitation of existing overlays is limited support for control logic. Most existing overlays instead primarily focus on computationally intensive datapath logic using pipeline-centric \cite{coole:adjustable} and/or processor-centric overlays \cite{li:area}. Although datapath logic is generally responsible for lengthy compile times and the majority of a design's resources, by themselves, datapath-centric overlays do not provide a significant improvement to productivity for realistic use cases. Although creating a specialized datapath may initially require more time than a controller, based on our observations, the controller is much more likely to change during development (e.g., iterative design, testing, and debugging) \cite{sklyarov:reconfigurable}. Control overlays can therefore provide the design flexibility to meet unpredictable requirements during these situations. 

Therefore, for productivity improvements to be realized in common design methodologies, existing overlays need to be complemented by appropriate control overlays that allow for rapid changes in control without full-detail FPGA compiles.

Several existing approaches have introduced such overlays, but those architectures have limited scalability for large applications and changing control requirements \cite{cooke:fsm}. Similarly, those overlays support only single FSM applications, which can be impractical for parallel applications that experience ``state explosion" in their single FSM implementations \cite{clarke:progress}. 

In this paper, we introduce a new control overlay---the Multi-RAM architecture (M-RAM for short)---that addresses these limitations. Using memory decomposition, this overlay enables enhanced scalability and flexibility, while also supporting parallel finite-state machines (FSMs). As shown in Figure \ref{fig:overview}, one envisioned use for this new control overlay is to complement existing datapath overlays in a high-level synthesis approach that compiles high-level code to an overlay instance tailored to an application's requirements (e.g., \cite{coole:adjustable}). Such an approach enables high productivity via the use of high-level code, and fast compilation for rapid iterative development methodologies, while also enabling portability across different FPGAs. For the common use case of supporting multiple FSMs in a single instance, the M-RAM achieves a 77\% to 99\% reduction of Virtex 7 lookup tables compared to previous architectures. On average, the M-RAM has an average clock frequency of 203 MHz and reduces compilation times to tenths of a second.

\section{Related Work} \label{sec:related}

Existing overlay research has focused largely on identifying appropriate tradeoffs between resource specialization, flexibility, and overhead. An overly specialized overlay may have low overhead, but will not improve productivity due to the need to fallback on full-detail FPGA compilation to support a changing application. Similarly, an overly general overlay may be flexible, but may incur prohibitive overheads \cite{li:area}.

Most existing overlays have focused on applying these tradeoffs on pipelined datapaths with minimal control requirements through the interconnection structure (e.g.,\cite{coole:intermediate,coole:adjustable,jain:deco,shukla:quku}). Previous work in control-specialized overlays primarily focused on memory-based FSM implementations \cite{cooke:fsm} using general techniques for FSM synthesis \cite{micheli:synthesis} and decomposition techniques from reconfigurable FSM studies (e.g.,\cite{borowik:statechart,vargas:fsm,glaser:fsm}). The Multi-RAM overlay expands from the 3-RAM overlay \cite{cooke:fsm} with memory decomposition to reduce memory requirements and comparatively requires 15\% to 28\% fewer lookup tables for individual FSMs, and 77\% to 99\% fewer lookup tables for more common overlay use cases.

Vendor tools have introduced different techniques to minimize compilation time, such as partial reconfiguration (PR) \cite{xilinx:partial} and incremental compilation \cite{altera:increasing}. Although both techniques may grant modest reductions in  compile time, both notably require additional effort on the designer and produce device-specific files. For an overlay implementation, the FSM controller is compiled to an overlay bitstream that is portable to any FPGA that supports that overlay instance.

\section{Background} \label{sec:backgnd}
This section describes three FSM overlays from previous works \cite{cooke:fsm}, which we extend with the M-RAM overlay. 

The 1-RAM architecture is a common reconfigurable FSM architecture that consists of a state-transition RAM and a state register. Using the current state from the state register and input values, the state-transition RAM acts as a lookup table to find the corresponding next state and respective output values. 

The 2-RAM architecture is conceptually similar to a number of works \cite{vargas:fsm}\cite{sklyarov:reconfigurable}. The key difference from the 1-RAM is the use of \emph{effective inputs}, which are the subset of total inputs that affect a state transition in a particular state. Ideally, the state-transition RAM should only store input combinations for inputs that can cause a transition. Compared to the 1-RAM, the 2-RAM architecture also uses an input-selection RAM and a series of input multiplexers to select the effective inputs of the current state, given by the state register. Using these effective inputs and the current state, the state-transition RAM looks up the corresponding next state and respective output values. 

The 3-RAM architecture \cite{cooke:fsm} extends from the 2-RAM with transition references. The main source of overhead in the 2-RAM is that the state-transition RAM stores transition data for every combination of inputs, even for the same transition. Ideally, the state-transition RAM could instead store references to a transition, which avoids redundant replications of the transition's output.
Compared to the 2-RAM, the 3-RAM architecture uses the state-transition RAM to store a transition index rather than transition outputs. Using this transition index and the current state, a new transition-code RAM looks up the respective next state and output values. 

Both the 2-RAM and 3-RAM are notably limited when the maximum number of effective inputs that governs the size of the state-transition RAM is much larger than the number of effective inputs across a majority of the FSM's states. 

\section{Multi-RAM Architecture} \label{sec:arch}
In this section, we present the new M-RAM architecture, which extends from the 3-RAM architecture \cite{cooke:fsm} with the use of memory decomposition on the state-transition RAM to avoid excessive overhead when an FSM has a wide range in the amount of effective inputs across its states.  

\subsection{Motivation} \label{subsec:problem}
Previous FSM overlays implement transition logic using memory as a lookup table. Mapping transition logic to memory is conceptually similar to mapping an $n$-input function to a $m$-input lookup table, where $m \geq n$. When the function and the lookup table have the same number of inputs ($m=n$), the lookup table encodes the exact truth table of the function. A major source of redundancy occurs when the lookup table has a larger number of inputs than the function ($m>n$). In this scenario, the excess inputs of the lookup table have no effect on the output of the function (i.e., don't care inputs), but the encoded truth table must produce the correct values for any combination of excess input values. As such, the encoded truth table resembles the function's truth table replicated for each combination of excess input values.

\begin{figure}[ht]
\centering
	\includegraphics[width=0.8\columnwidth]{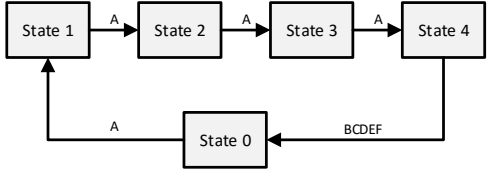} \\
\caption{Example of a simple 5-state FSM}
\label{fig:example}
\end{figure}

As previously mentioned, the 2-RAM and 3-RAM use ``effective inputs" in the state-transition RAM to implement transition logic. In these overlays, the state-transition RAM is sized by a maximum effective input and encodes transition functions for all states, even those with a vastly smaller number of effective inputs.

\begin{figure}[h]
\centering
\includegraphics[width=0.9\columnwidth]{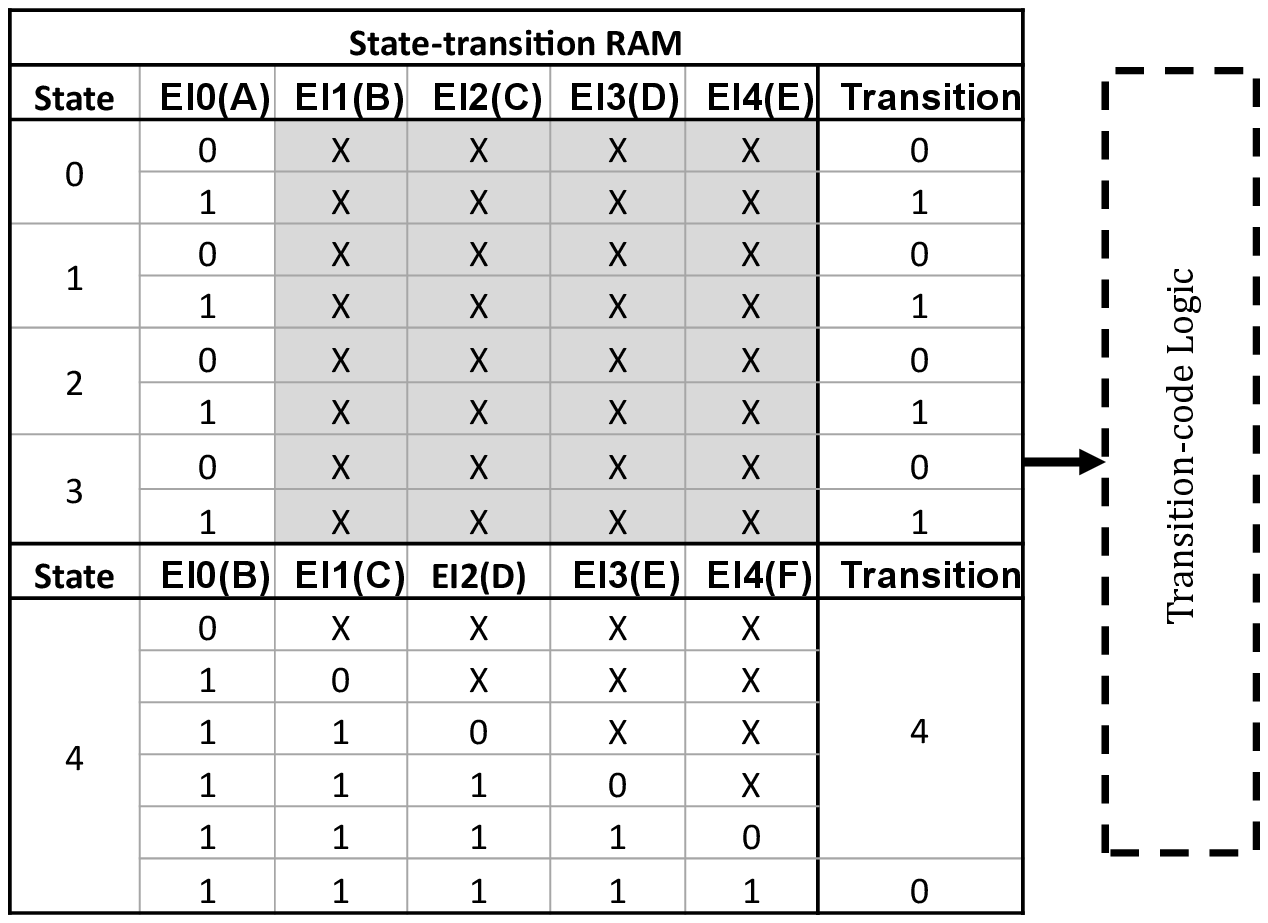}
\caption{FSM mapping on 3-RAM's state-transition RAM}
\label{fig:mapping_3ram}
\end{figure}

To better illustrate this problem, consider the simple 5-state FSM in Figure \ref{fig:example}. In states 0 to 3, the FSM transitions to the next numbered state when input A is true. In state 4, the FSM transitions to state 0 when inputs B, C, D, E, and F are true. The maximum number of effective inputs is five since states 0 to 3 have one effective input (A), and state 4 has five effective inputs (B, C, D, E, and F). 
For the previous overlays, the state-transition RAM must encode each state's transition function in an array of five-input truth tables as shown in Figure \ref{fig:mapping_3ram}, where five is the max number of effective inputs (EI) across all states. For the four states with a single effective input, their transition function is encoded for every combination of the four excess inputs, yielding $2^4$ copies, even though only one state has the same effective input number as the maximum. In Section \ref{sec:experiments}, we show that these replications can become a significant source of overhead.

\subsection{Architecture} \label{subsec:arch}

The M-RAM architecture addresses the unnecessary replication described in the previous section by extending the 3-RAM with memory decomposition on the state-transition RAM. By using a collection of smaller-sized RAMs targeting different numbers of effective inputs, the M-RAM can map each state to a memory that encodes the state's truth table without replication. The M-RAM can therefore avoid a large source of replication seen in larger FSMs.

\begin{figure}[ht]
\centering
\includegraphics[width=\columnwidth]{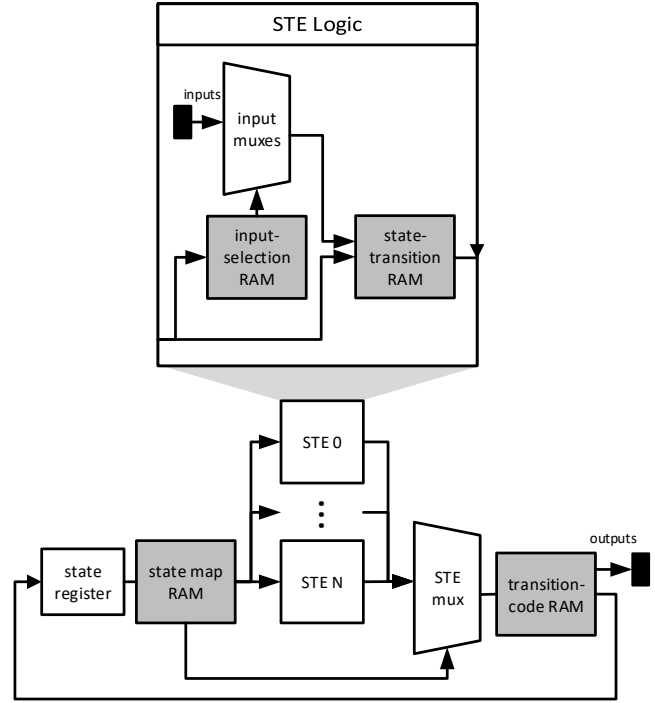}
\caption{Multi-RAM Architecture}
\label{fig:mram}
\end{figure}

In Figure \ref{fig:mram}, we illustrate a high-level block diagram of the M-RAM. The main difference from the 3-RAM is the decomposed memory elements which implement the state transition functions, which we refer to as state-transition elements (STEs). Each STE is a grouping of input muxes and smaller-sized input-selection and state-transition RAMs targeting a different effective input number. Using these elements, the architecture maps each FSM state to a \emph{pseudo state} in a STE when that particular STE can directly encode the state's transition function without replication. Similar to the state variable in the 3-RAM's state-transition RAM, each STE has a designated number of pseudo states that index distinct state transition functions. Unlike previous architectures, the state-transition RAM stores transition indexes relative to the entire FSM rather than on a state-by-state basis. The state-transition RAM's size in a STE $i$ is:
\begin{equation}
2^{\lceil log_{2}(S_{i,total})\rceil +EI_{i}}*\lceil log_{2}(T_{max})\rceil
\end{equation}
whereas the input-selection RAM's size in a STE $i$ is:
\begin{equation}
2^{\lceil log_{2}(S_{i,total}) \rceil}*EI_{i}*\lceil log_{2}(I_{total})\rceil 
\end{equation}
where $S_{i,total}$ corresponds to the total number of pseudo states supported by the STE $i$, $EI_{i}$ corresponds to the effective input number targeted by the STE $i$, $T_{max}$ corresponds to the maximum number of transitions in the FSM, and $I_{total}$ corresponds to the total number of inputs in the FSM.

To use these decomposed elements, the architecture adds another layer of indirection before the state-transition logic through the state-map RAM. The state-map RAM stores mappings of FSM states to STE indexes and STE pseudo-states which allows the architecture to implement state transition functions across different STEs. The state-map RAM's required memory bits are as follows:
\begin{equation}
2^{\lceil log_{2}(S_{total})\rceil}*(\lceil log_{2}(S_{STE,max})\rceil+\lceil log_{2}(num_{STE})\rceil)
\end{equation}
where $S_{total}$ is the number of FSM states, $S_{STE,max}$ is the max number of pseudo-states across all STEs, and $num_{STE}$ is the number of STEs.

Using the current STE index from the state-map RAM, the STE mux directs the correct STE output to the transition-code RAM. The transition-code RAM then looks up the transition's respective next state and output values. Since the transition indexes no longer pertain to a state-by-state basis, the transition-code RAM's size in bits is as follows:
\begin{equation}
2^{\lceil log_{2}(T_{max})\rceil}*(\lceil log_{2}(S_{total})\rceil+O_{total})
\end{equation}

\begin{figure}[h]
\centering
\includegraphics[width=\columnwidth]{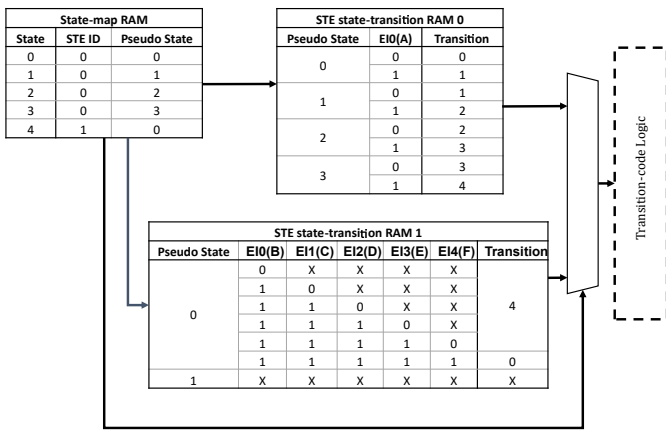}
\caption{FSM mapping on M-RAM architecture}
\label{fig:mapping_mram}
\end{figure}

To better illustrate the advantage of this architecture, consider the same example 5-state FSM discussed in Section \ref{subsec:problem}. In Figure \ref{fig:mapping_mram}, we illustrate the FSM mapping on the M-RAM architecture. Unlike the 3-RAM, the transition functions of states with single effective inputs are encoded directly onto STE 0's state-transition RAM at pseudo states 0 to 3, and the transition function of the single state with 5 effective inputs is directly encoded onto STE 1's state-transition RAM at pseudo state 0. Similar to the 3-RAM's mapping, not all possible state values are used in the binary encoding, leaving STE 1's pseudo state 1 unused. By mapping these states to different STEs in the state-map RAM, the architecture has avoided redundant replications due to states using fewer than the max number of effective inputs. This simple example shows a M-RAM architecture instance requiring 306 RAM bits, which is a 27\% reduction compared to the 3-RAM architecture's 424 RAM bits. 

Similar to past FSM overlays, the M-RAM overlay only maintains these advantages for scenarios that benefit its decomposed structure, which are large FSMs with a wide range of effective input numbers. Other usage scenarios favor past overlays, such as very small FSMs for 1-RAM, larger FSMs with small numbers of outputs for 2-RAM, and larger FSMs with a small range of effective input numbers for 3-RAM.

\subsection{Implementation} \label{subsec:impl}
The M-RAM is parameterized by the numbers of states, unique FSM transitions, inputs, outputs, and STEs, as well as each STE's number of states and effective inputs. Although an instance is limited to supporting FSMs with compatible numbers of states, transitions, inputs, and outputs, there is flexibility in the FSM mapping on STEs during overlay reconfiguration, where a STE with a larger supported effective input number may map states with a lower effective input number through redundant replications. Although past overlays used fewer yet more flexible parameters, the significant area savings in the Multi-RAM may be used to increase the size of the M-RAM parameters to significantly improve flexibility.

Conveniently, using multiple STEs grants opportunities for mapping parallel FSMs onto unused STEs in a single overlay instance. Whereas previous FSM overlays must replicate the architecture for each parallel FSM, the M-RAM architecture can add support by replicating the non-STE components for each FSM and by adding additional glue logic. In these instances, the glue logic will act as an overlay-configured interconnect that will ``assign" STEs and outputs to specific FSMs through overlay configuration bits. Since the STEs can be tailored to more efficiently fit an FSM's transition logic than a single memory, an appropriately configured M-RAM instance may support parallel FSMs at a much lower area overhead than previous FSM overlays. 
We plan to explore this feature as future work.

\subsection{FSM Mapping} \label{subsec:mapping}
A compatible FSM can be mapped to an overlay instance by writing the appropriate memory contents to the RAM structures. Due to the overlay's decomposed nature, the FSM must be similarly decomposed. First, the mapper creates a list of unique transitions (next state and output value pairs) which will be written to the transition-code RAM. Second, the mapper assigns each state to a STE pseudo-state targeting an effective input number equal to or greater than the state's effective input number. From these assignments, the mapper will create a mapping of pseudo-state and every combination of STE input values to a transition-code RAM address that matches the assigned state's transition outputs, which will be written to the STE's state-transition RAMs. Similarly, the mapper will create a mapping of pseudo-state to the assigned state's effective input indices, which will be written to the STE's input-selection RAMs. Finally, using the previous state assignment, the mapper creates a mapping of state to pseudo-state and STE IDs, which will written to the state-map RAM.

Due to the complexity of mapping FSMs, the M-RAM would ideally be complemented with a high-level synthesis process that automatically handles both implementation and mapping, which we will investigate as future work. 

\begin{table}[ht]
  \small
  \centering
  \caption{Area comparison of FSM overlays and direct RTL}
  	\setlength\tabcolsep{2pt}
    \begin{tabular}{c|cccc|cc}
    \toprule
		\multirow{3}{*}{Benchmark} & \multicolumn{4}{c|}{\multirow{2}{*}{Virtex 7 LUTs}}  & \multicolumn{2}{c}{M-RAM LUT}\\
		& & & & & \multicolumn{2}{c}{Reduction} \\
		& M-RAM & 3-RAM & 2-RAM & RTL & 3-RAM & 2-RAM \\
    \midrule
    origin  & 12    & 9     & 7     & 1     & -33\% & -71\% \\
    s298    & 797   & 926   & 759   & 157   & 14\%  & -5\% \\
    opus    & 68    & 93    & 119   & 23    & 27\%  & 43\% \\
    styr    & 382   & 507   & 2K    & 90    & 25\%  & 76\% \\
    sync    & 93    & 75    & 65    & 45    & -24\% & -43\% \\
    s510    & 121   & 120   & 98    & 53    & -1\%  & -23\% \\
    s208    & 112   & 80    & 97    & 12    & -40\% & -15\% \\
    s420    & 134   & 94    & 117   & 12    & -43\% & -15\% \\
    ex2     & 41    & 47    & 35    & 12    & 13\%  & -17\% \\
    ex1     & 379   & 415   & 861   & 63    & 9\%   & 56\% \\
    sand    & 413   & 767   & 2K    & 94    & 46\%  & 75\% \\
    s832    & 732   & 1K    & 4K    & 77    & 38\%  & 83\% \\
    s820    & 640   & 1K    & 4K    & 84    & 46\%  & 85\% \\
    kirkman & 9K    & 9K    & 24K   & 24    & 2\%   & 63\% \\
    s1494   & 420   & 2K    & 2K    & 123   & 73\%  & 80\% \\
    scf     & 1K    & 8K    & 136K  & 164   & 88\%  & 99\% \\
    \midrule
    multi1  & 1K    & 12K   & 136K  & N/A   & 89\%  & 99\% \\
    multi2  & 1K    & 8K    & 10K   & N/A   & 83\%  & 87\% \\
    multi3  & 8K    & 36K   & 183K  & N/A   & 77\%  & 95\% \\
    all     & 10K   & 222K* & 2M*   & N/A   & 95\%  & 99\% \\
    \midrule
    Bench. Avg.    & 876   & 1K    & 11K   & 65    & 15\%  & 29\% \\
    Bench. Median  & 381   & 461   & 810   & 58    & 13\%  & 49\% \\
    Trimmed Avg. & 356   & 1K    & 10K   & 67    & 16\%  & 27\% \\
    Total Avg. & 2K    & 15K   & 125K  & 65    & 29\%  & 43\% \\
    \bottomrule
    \multicolumn{7}{c}{Key: K = $\times 10^3$, M = $\times 10^6$} \\
    \multicolumn{7}{c}{* example does not fit on XC7VX485T} \\
    \end{tabular}%
  \label{tab:area_async}%
\end{table}%

\section{Experiments} \label{sec:experiments}
The following subsections define the experimental setup (Section \ref{subsec:setup}), and evaluate the area (Section \ref{subsec:area}), clock (section \ref{subsec:clk}), and compilation time (Section \ref{subsec:time}) of the M-RAM compared to existing FSM overlays and direct RTL.

\subsection{Experimental Setup} \label{subsec:setup}

For each experiment, we evaluate the different architectures using FSMs from the IWLS 93 benchmark set \cite{iwls93}. Since the overhead of overlay architectures described in Section \ref{sec:backgnd} is largely determined by $I_{total}$ and $S_{total}$, we selected 16 benchmarks to cover a wide range of $I_{total}$ and $S_{total}$. 

In Sections \ref{subsec:area} and \ref{subsec:clk}, we analyze the area and clock of the M-RAM compared to existing overlays and direct RTL. For these experiments, we obtain synthesis results of the overlay architectures and the FSM's direct RTL implementations using Vivado 2015.2 targeting the XC7VX485T FPGA. 

In Section \ref{subsec:time}, we analyze compile times for FSMs on the M-RAM architecture, on existing overlays, and directly on the FPGA. We use Python 2.7 scripts to generate overlay bitstreams from the KISS format \cite{iwls93}. For the direct RTL implementations, we use Vivado 2015.2 to generate the FPGA bitstream for the same KISS FSMs converted to Verilog  \cite{pruteanu:converter}.

\subsection{Area Analysis} \label{subsec:area}
This experiment focuses on comparing the area of the M-RAM overlay with prior FSM-based overlays and with direct RTL. For each benchmark, the overlay instance is tailored to minimally support the benchmark. For our synthetic examples, the overlay instance is tailored to minimally support a subset of the benchmark set. It should be noted that these instances still support any FSM whose parameters do not exceed an overlay's limits. From here on, any references to lookup tables (LUTs) refer to the FPGA's physical fine-grained LUTs, as opposed to the logical LUTs implementing functions as described in Section \ref{sec:arch}, which we implement using distributed RAM.

Table \ref{tab:area_async} shows the LUT results for the overlays and for direct RTL. The ``Virtex 7 LUTs" section reports the number of Virtex 7 LUTs after synthesis, translation, and mapping with post-synthesis estimates for designs that do not fit. As a point of reference, we provide synthesis results for direct RTL, whose implementations are often significantly smaller than respective overlay implementations due to gate-level optimizations. The ``M-RAM LUT Reduction" section reports the M-RAM's reduction in LUTs relative to the 3-RAM and 2-RAM. For each section, the table shows the average and median of results for individual benchmarks, a trimmed average excluding the $kirkman$ as an extreme outlier for all overlays, and a total average including synthetic multi-FSM examples.

To evaluate the scalability of the overlays, we tested a synthetic example (shown as $all$) that requires support for overlay reconfiguration of each benchmark from the benchmark set in a single architecture instance. In this experiment, the M-RAM LUT reduction is 95\% compared to the 3-RAM and 99\% compared to the 2-RAM. Notably, the M-RAM is the only overlay that fits on the target Virtex 7 device with a 3.31\% LUT utilization, whereas the 3-RAM and 2-RAM require more than the total amount of available memory LUTs. Similarly, we tested additional examples (shown as $multi($1-$3)$) that require support for overlay reconfiguration of each benchmark from a subset of randomly selected benchmarks and found the M-RAM demonstrates similar trends with average LUT reductions of 83\% for the 3-RAM and 94\% for the 2-RAM. These results for the synthetic examples are the most accurate reflection of typical use cases, where the overlays must handle flexible control requirements rather than minimally support a specific application. Overall, the M-RAM was the only overlay that could support such control variability.

For individual benchmarks, the M-RAM shows average LUT reductions of 15\% compared to the 3-RAM and 29\% for the 2-RAM. Note the average compared to the 3-RAM is skewed due to a high range of reduction values that result from differently structured benchmarks, especially the smaller benchmarks which represent minimal flexibility.
As such, the major trend is that the M-RAM performs most favorably for larger benchmarks ($s820$, $s1393$, $scf$) and unfavorably for smaller benchmarks ($origin$, $s208$, $s420$), where the cost of decomposition is significant.

\subsection{Clock Analysis} \label{subsec:clk}
This experiment focuses on comparing the clock frequencies of the M-RAM and prior FSM-based overlays with direct RTL. Like prior experiments, the overlay instance for each benchmark is tailored to minimally support it, with the exception of the synthetic examples that support subsets of benchmarks.

\begin{figure}[h]
\centering
\includegraphics[width=\columnwidth]{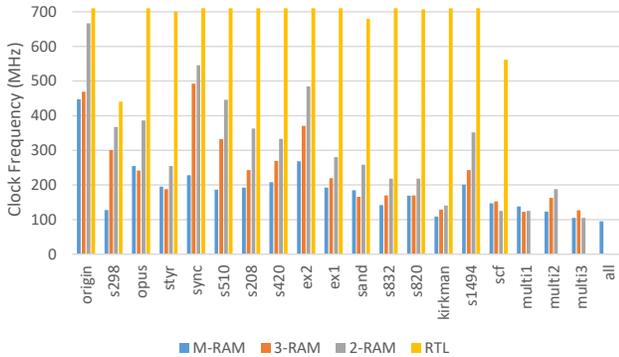}
    
\caption{Clock frequency comparison of overlays and direct RTL}
\label{fig:clk_async}
\end{figure}

Figure \ref{fig:clk_async} shows the clock frequency results for each overlay with a direct RTL. Since these designs produce same-cycle outputs, each overlay has increasing clock overhead for each additional RAM in the critical path, with the M-RAM having the most RAMs with the state-map RAM addition. For individual benchmarks, we find clock overheads compared to direct RTL of 70\% for the M-RAM, 61\% for the 3-RAM, and 50\% for the 2-RAM. Despite having the largest clock overhead, the M-RAM is still able to provide an average clock frequency of 203 MHz, which is unlikely to be prohibitive for many applications where datapath elements typically operate at lower clock frequencies. 

For applications that tolerate multi-cycle control outputs, FSM overlays can implement RAMs using block RAM instead of distributed RAM. For these cases, the M-RAM has a greater average clock frequency of 392 MHz. Due to the abundance of block RAM for various devices, ideally the M-RAM overlay would be implemented with a mix of distributed RAM and block RAM to minimize area and clock overhead.

\subsection{Compilation Time} \label{subsec:time}
This experiment compares FSM compile times of overlays with Vivado 2015.2. For the different overlays, the M-RAM has the shortest average compilation time at 0.236 seconds, which is a 76\% reduction compared to the 3-RAM's 0.999 seconds, and the 2-RAM's 1.012 seconds. Notably the M-RAM's compilation time is faster on the larger benchmarks due to its smaller memory imprint and overlay bitstream size. 

All overlays achieve four orders-of-magnitude average improvements compared to Vivado's average of 553 seconds. As FPGA devices increase the number of resources, these improvements will increase as overlay instances remain constant across devices. 

\section{Conclusion} \label{sec:conclusions}
In this paper, we introduced a new FSM-based overlay architecture that complements existing datapath-centric overlays to enable rapid compilation for iterative application development. The presented Multi-RAM overlay uses memory decomposition on an FSM's transition logic to efficiently map a state transition graph's representation onto a collection of smaller memories, rather than on a single large memory. Although showing modest improvements for individual FSMs, compared to previous work, this architecture enables enhanced scalability and flexibility showing a 77\% to 99\% reduction in Virtex 7 lookup tables when supporting multiple benchmarks in a single overlay instance. Similarly, the architecture reduces finite-state machine compilation times to tenths of a second and enables potential support for parallel FSMs. For future work, we intend to explore the integration of control and datapath overlays.

\section*{Acknowledgement}
This work was supported by the I/UCRC Program of the National Science Foundation under Grant No. EEC-0642422 and IIP-1161022.

\bibliographystyle{abbrv}

\end{document}